\documentclass[sigconf]{acmart}
\AtBeginDocument{%
  }

\setcopyright{acmlicensed}
\copyrightyear{2018}
\acmYear{2018}
\acmDOI{XXXXXXX.XXXXXXX}
\acmConference[CHI '25 Indigeneity and Culture Workshop]{Weaving Indigeneity and Culture into the Fabric of HCI Futures: CHI '25 Workshop}{April 26th, 2025}{Yokohama, Japan}

\begin{document}

\title{Speculative Design in Spiraling Time: Methods and Indigenous HCI}

\author{James Eschrich}
\email{jamesae2@illinois.edu}
\orcid{0009-0009-7608-3862}
\affiliation{%
  \institution{University of Illinois, Urbana-Champaign}
  \city{Urbana}
  \state{Illinois}
  \country{USA}
}
\author{Cole McMullen}
\email{cdm45@uw.edu}
\affiliation{%
  \institution{University of Washington}
  \city{Seattle}
  \state{Washington}
  \country{USA}
}

\author{Sarah Sterman}
\email{ssterman@illinois.edu}
\orcid{0000-0002-9282-559X}
\affiliation{%
  \institution{University of Illinois, Urbana-Champaign}
  \city{Urbana}
  \country{Illinois}}
\renewcommand{\shortauthors}{Eschrich et al.}
\begin{abstract}
  In this position paper, we first discuss the uptake of speculative design as a method for Indigenous HCI. Then, we outline how a key assumption about temporality threatens to undermine the usefulness of speculative design in this context. Finally, we briefly sketch out a possible alternative understanding of speculative design, based on the concept of "spiraling time," which could be better suited for Indigenous HCI.
\end{abstract}

\ccsdesc[500]{Do Not Use This Code~Generate the Correct Terms for Your Paper}
\ccsdesc[300]{Do Not Use This Code~Generate the Correct Terms for Your Paper}
\ccsdesc{Do Not Use This Code~Generate the Correct Terms for Your Paper}
\ccsdesc[100]{Do Not Use This Code~Generate the Correct Terms for Your Paper}

\keywords{Speculative Design, Indigenous speculation, design futures}



\maketitle

\section{Introduction}
Speculative design as a research method has been used to make space for subaltern voices in HCI. In some cases, the research team performs and reflects on the process of speculation \cite{sondergaardFabulationApproachDesign2023,klassenStoopSpeculationPositive2022}; in other cases, speculative and participatory design methods are combined \cite{fariasSocialDreamingTogether2022}, and participants from marginalized communities work with the research team to develop speculative visions of alternative futures \cite{brayRadicalFuturesSupporting2022, khanSpeculativeDesignEducation2021, zolyomiEmotionTranslatorSpeculative2024, klassenBlackFuturePower2024}. Indigenous researchers and researchers working with indigenous people are no exception to this rule; examples include \cite{muashekeleRecoveringLostFutures2024, muashekeleAncestralCulturalFuturing2023, akamaSpeculativeDesignHeterogeneity2016}.

Kyle Whyte's work \cite{whyteIndigenousScienceFiction2018} suggests two reasons Indigenous scholarship in HCI might take up speculative design as a method. The first reason is that the ``dramatically curtailed collective agency'' of Indigenous societies today as the result of settler-colonialism represents ``a situation that [their] ancestors ... would have seen as a dystopian science fiction scenario'' \cite[p.~228]{whyteIndigenousScienceFiction2018}. Conversely, from the perspective of settler-colonialist ancestors, the present, in which "their descendants ... have the privilege of unlimited individual and collective agency to exploit Indigenous peoples \textit{and} the privilege of claiming moral high ground as saviors" \cite[p.~238]{whyteIndigenousScienceFiction2018}, represents a kind of utopian fantasy. As a result, advancing Indigenous critiques of or centering Indigenous perspectives on the present often demands the language of speculative and/or science fiction. Speculative design as a method helps legitimize that language in HCI research.

The second reason Indigenous scholarship in HCI might turn to speculative design has to do with Indigenous narrative and epistemic practices. These practices, Whyte argues, rely on different models of temporality \cite{whyteIndigenousScienceFiction2018} than the linear, progressional model assumed by designers and HCI researchers. James Pierce's analysis of speculative design as "in tension with progression" \cite{pierceTensionProgressionGrasping2021} suggests that research which uses design methods and centers Indigenous narrative and epistemic practices might almost inevitably become classified as "speculative design" within the HCI community.

We find speculative design to be an exciting and generative research method ourselves, and are hopeful that it will continue to provide space within HCI for Indigenous and other marginalized perspectives. However, we are also concerned that the temporal logic used to theorize about, structure, and communicate speculative design projects is problematic for this purpose. In this position paper, we will outline these issues and consider how the concept of Indigenous "spiraling time" \cite{whyteIndigenousScienceFiction2018} could help develop new understandings of speculative design as a research method.

\section{Time, Design, Capital}
As Martijn Konings argues in \textit{Capital and Time: For a New Critique of Neoliberal Reason} the “rise of modern capitalism was accompanied by the emergence of a secular experience of time." People began "understanding present practices as having emerged out of a past and as shaping a contingent future;" in other words, we began to understand ourselves as "making history" [p.~71]. Time became "a source of variation" and the future "open, its shape not yet determined or known" [p.~72]. 

\begin{figure}
    \centering
    \includegraphics[width=0.8\linewidth]{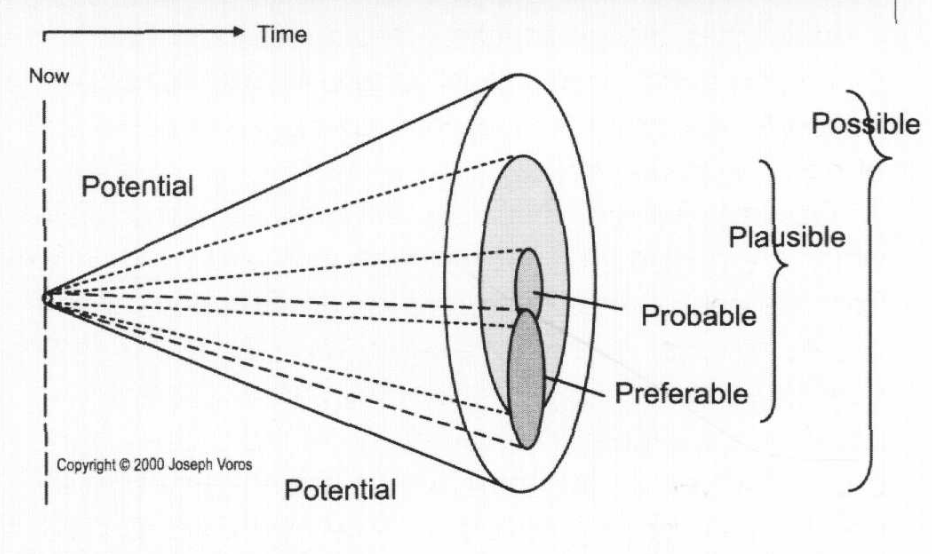}
    \caption{An illustration of the "futures cone," taken from \cite{vorosGenericForesightProcess2003}}
    \label{fig:cone}
\end{figure}

As the modern discipline of design emerged during the industrial revolution \cite[p.~3]{pierceTensionProgressionGrasping2021}, it is no surprise theoretical analyses of design—and, thus, speculative design—inherit this understanding of time. This understanding is often represented via a diagram known as the "futures cone" \cite{vorosGenericForesightProcess2003} (see Figure \ref{fig:cone}). In this diagram, the variable and contingent future is represented as bounded, cone-shaped possibility space; smaller cones identify sections of that space which are more or less likely and more or less desirable. In James Pierce's analysis \cite{pierceTensionProgressionGrasping2021}, design is the process of creating "partial, provisional, and potentially preliminary material actualizations of possible futures" \cite[p.~6]{pierceTensionProgressionGrasping2021}; in essence, explorations of the futures cone. Conventional design, under Pierce's theory, produces designs which "ultimately tend toward convergence toward production" \cite[p.~7]{{pierceTensionProgressionGrasping2021}}; in contrast, speculative design produces designs which "appear deliberately and intriguingly resistant towards production" \cite[p.~9]{pierceTensionProgressionGrasping2021}. 

Under Pierce's model both conventional and speculative designs derive their value from two factors: how possible and how desirable the future they prefigure is. In this way, Pierce's model is an excellent model of design under modern capitalism: as Konings argues, assigning value to speculations along these axes is an essential part of the process by which capitalism reproduces and perpetuates itself \cite{koningsCapitalTimeNew2020}. However, does this mean that speculative design inevitably replicates the temporal logic of modern capitalism? If so, is the use of speculative design as a research method for Indigenous HCI appropriate? If not, what might an alternative approach to speculative design, based on an alternative model of temporality, look like? 

\section{Spiraling Time and Speculative Design}
In this section, we sketch an alternative approach to speculative design, one which is not based on the temporal logic of modern capitalism, but instead on the Indigenous concept of "spiraling time." As Whyte defines it, spiraling time is a perspective on time in which Indigenous people consider themselves "as living alongside future and past relatives simultaneously" \cite[p.~229]{whyteIndigenousScienceFiction2018}. While this does not "foreclose linear, future thinking" \cite[p.~229]{whyteIndigenousScienceFiction2018} it does decenter and disrupt it. Understanding time as "a dialogical unfolding" with a kind of "forward motion" which "can be both predictable and irregular," spiraling time makes room for "narratives of cyclicality, reversal, dream-like scenarios, simultaneity, counter-factuality, irregular rhythms, ironic un-cyclicality, slipstream, parodies of linear pragmatism, eternality, among many others" \cite[p.~229] {whyteIndigenousScienceFiction2018}. Critically, "the dialogical unfolding of spiraling time" also supports "a form of philosophizing" through "counterfactual dialogue" in which speculation occurs about how past and futures "generations would interpret today's situations and what recommendations they would make" \cite[p.~229]{whyteIndigenousScienceFiction2018}. 

Spiraling time, and the notion of counterfactual dialogue which accompanies it, could represent an alternative framework for speculative design. Rather than understanding speculative design as an exploration or prefiguration of possible futures, speculative design could be understood as a method for staging counterfactual dialogues with ancestors and descendants. The value of a speculative design, understood through the lens of spiraling time, would not rely on its relationship to "possibility" or "desirability," but instead on its intrinsic value as a counterfactual representation of the "future and past relatives" alongside which people "walk through life" \cite[p.~229]{whyteIndigenousScienceFiction2018}. In this way, spiraling time could enable speculative design to be understood, practiced, and communicated without reference to the temporal logic of modern capitalism.

As an example of what this could look like in practice, consider Muashekele et al. \cite{muashekeleAncestralCulturalFuturing2023}. While they do not employ the notion of spiraling time in their study, their reflections on the use of participatory speculative design with Indigenous communities anticipates many of the points made in this section. They propose "positioning speculative design as a space for negotiation and reflection" \cite[p.~94]{muashekeleAncestralCulturalFuturing2023} in Indigenous contexts, and note that the communities they worked with "indicated the prominence and influence of the ancestors on the future" \cite[p.~93]{muashekeleAncestralCulturalFuturing2023}. Muashekele et al. also note that their participants were skeptical about "the speculation," highlighting frictions that emerged from applying a conventional understanding of speculative design in Indigenous contexts \cite[p.~93]{muashekeleAncestralCulturalFuturing2023}. A spiraling time approach to speculative design might have reduced some of these frictions. For example, Muashekele et al. describe intergenerational dialogue among their participants \cite[p.~92]{muashekeleAncestralCulturalFuturing2023}. These conversations are interpreted as failed attempts at conventional speculation in \cite{muashekeleAncestralCulturalFuturing2023}; a spiraling time perspective might have rendered these conversations legible as a successful, epistemically rich form of Indigenous speculation instead \cite[p.~93]{muashekeleAncestralCulturalFuturing2023}.

\section{Positionality Statement and Conclusion}
It is important to note that we, the authors of this paper, are not Indigenous ourselves. While the notion of spiraling time is a conception of temporality common "across different [Indigenous] cultures" \cite[p.~229]{whyteIndigenousScienceFiction2018} and has been proposed by some scholars as one of the contributions that "traditional indigenous knowledge forms" might make "towards an alternative ontology for a just global order" \cite[p.~32]{stewart-harawiraNewImperialOrder2005}, this does not necessarily imply that the concept of spiraling time should or could be naively absorbed into speculative design practice or design theory. How and to what extent non-Indigenous scholars should use and build on Indigenous ontologies and epistemologies is a question we would be interested in discussing during the workshop. 

Ultimately, however, we feel that theories of design—speculative design in particular—should explicitly and intentionally engage with Indigenous thought and make space for Indigenous ontologies and epistemologies. In doing so, theories of speculative design can be developed which are not tied to the temporal logic of modern capitalism. We believe such theories could help further develop speculative design as a method and better support "social dreaming" \cite[p.~vi]{dunneSpeculativeEverythingDesign2013}.

\begin{acks}
Thanks to John Marshall for conversations about design theory.
\end{acks}

\bibliographystyle{ACM-Reference-Format} 
\bibliography{article}
\end{document}